\documentclass[a4paper,twoside]{article}

\usepackage{epsfig}
\usepackage{subcaption}
\usepackage{calc}
\usepackage{amssymb}
\usepackage{amstext}
\usepackage{amsmath}
\usepackage{amsthm}
\usepackage{multicol}
\usepackage{pslatex}
\usepackage{apalike}
\usepackage{algorithm2e}
\usepackage[bottom]{footmisc}
\usepackage{SCITEPRESS}     

\usepackage{xcolor}
\usepackage{url}
\newcommand{\footurl}[1]{\footnote{\url{#1}}}

\newenvironment{itquote}
  {\begin{quote}\itshape}
  {\end{quote}\ignorespacesafterend}

\begin{document}

\title{Using GPT to build a Project Management assistant for Jira environments}

\author{\authorname{
Joel Garcia-Escribano\sup{1},
Arkaitz Carbajo\sup{2}\orcidAuthor{0000-0003-0904-4627},
Mikel Ega\~na Aranguren\sup{3}\orcidAuthor{0000-0001-8081-1839},
and Unai Lopez-Novoa\sup{3}\orcidAuthor{0000-0002-2707-8946}
}
\affiliation{\sup{1}Accenture, Bilbao, Spain}
\affiliation{\sup{2}LKS Next GobTech, Bilbao, Spain}
\affiliation{\sup{3}University of the Basque Country UPV/EHU,  Bilbao, Spain}
\email{joel.b.garcia@accenture.com, arkaitz.carbajo@gobtech.lksnext.com, \{mikel.egana, unai.lopez\}@ehu.eus}
}

\keywords{Intelligent Decision Making, LLM, Jira, GPT, Prompt Engineering, Project Management.}

\abstract{In the domain of Project Management, the sheer volume of data is a challenge that project managers continually have to deal with. Effectively steering projects from inception to completion requires handling of diverse information streams, including timelines, budgetary considerations, and task dependencies. To navigate this data-driven landscape with precision and agility, project managers must rely on efficient and sophisticated tools. These tools have become essential, as they enable project managers to streamline communication, optimize resource allocation, and make informed decisions in real-time. However, many of these tools have steep learning curves and require using complex programming languages to retrieve the exact data that project managers need. In this work we present JiraGPT Next, a software that uses the GPT Large Language Model to ease the process by which project managers deal with large amounts of data. It is conceived as an add-on for Jira, one of the most popular Project Management tools, and provides a natural language interface to retrieve information. This work presents the design decisions behind JiraGPT Next and an evaluation of the accuracy of GPT in this context, including the effects of providing different prompts to complete a particular task.}

\onecolumn \maketitle \normalsize \setcounter{footnote}{0} \vfill

\section{Introduction} \label{sec:intro} 

In the evolving landscape of Project Management, the confluence of artificial intelligence and task management platforms has become a focal point for driving efficiency and enhancing decision-making processes \cite{9610711,app13085014}. Jira\footurl{https://www.atlassian.com/software/jira}, developed by Atlassian, stands out as an industry-standard tool that enables teams to plan, track, and manage agile software development projects. However, despite its comprehensive feature set, Jira users often encounter challenges in navigating complex workflows, managing growing backlogs, and maintaining clear communication across tasks and team members. 

As the velocity of software development accelerates and the volume of tasks within projects proliferates, the need for intelligent automation within Project Management tools becomes increasingly paramount. The deployment of AI in this domain has predominantly focused on predictive analytics for project outcomes or robotic process automation for task completion. However, these applications have not fully leveraged the potential of AI to interact with users in a natural and contextually relevant manner. The advent of sophisticated language models such as GPT \cite{NEURIPS2020_1457c0d6} offers an untapped opportunity to address this gap  \cite{10.56889/nrkr7690}. 

This paper presents JiraGPT Next, a prototype of a GPT-based intelligent assistant for Jira environments. The aim is to provide a natural language interface by which managers can retrieve precise information about their projects in a seamless manner, hiding the complexities of Jira. Our software retrieves information from Jira database through its API and uses OpenAI's public API to access the available GPT models. 

JiraGPT Next has been developed as part of a collaboration with LKS Next-GobTech\footurl{https://www.lksnext.com/}, a division of the LKS Next Group focused on providing innovative solutions to improve the workflows of public administrations in their digital transformation. Some decisions about the design and development of the software have been taken with the constraints of this division in mind. Also, we want to note that, since LKS Next is based in Spain, this prototype has been implemented in Spanish. Translations to English are provided as necessary throughout this paper.

The rest of the paper is organised as follows: Section \ref{sec:background} describes the context on LLMs (\ref{sec:LLMs}) and Project Management (PM) \ref{sec:PM}; Section \ref{sec:JiraGPT Next} describes JiraGPT Next in detail, including the integration with GPT; Section \ref{sec:evaluation} provides an evaluation of the effects of using different GPT prompts for a particular task; Section \ref{sec:related-products} presents other tools applying AI to improve Project Management and a comparison of them with JiraGPT Next and finally, Section \ref{sec:concl} covers the wrapping conclusions and lies lines of future work.

\section{Background}\label{sec:background}

This section provides a description of Large Language Models (LLMs), the main AI tool behind the natural language replies in JiraGPT Next, and an overview of Project Management, the domain of application of this work.

\subsection{LLMs}\label{sec:LLMs}
LLMs have emerged as a transformative force in the realm of computer science, particularly in the domain of NLP (Natural Language Processing). Leveraging deep neural networks with millions, or even billions, of parameters, these models have been trained on extensive corpora of text data, enabling them to capture intricate patterns and relationships within natural language. LLMs are capable of learning contextual representations of words and phrases, resulting in a nuanced understanding of language semantics and syntax.

The architecture of LLMs is predominantly based on the transformer model \cite{vaswani2023attention}. The transformer model utilizes self-attention mechanisms to weigh the importance of different words in a sentence, enabling the capture of long-range dependencies and context. This architecture has proven to be highly effective for NLP tasks, leading to its widespread adoption in the development of LLMs.

The capabilities of LLMs have paved the way for groundbreaking applications across various fields, including but not limited to healthcare, finance, education, and customer service \cite{Author2023Prepare}. In healthcare, for example, LLMs assist in processing and summarizing patient records, enhancing the efficiency and accuracy of diagnostics and treatment planning. In the realm of customer service, they power chatbots and virtual assistants, providing quick and contextually relevant responses to user inquiries.

In many LLMs, including GPT, a prompt acts as the guiding instruction or query provided to the model to elicit a specific response. Crafting a prompt carefully is paramount, as it shapes the context and influences the nature of the generated output. A well-crafted prompt not only defines the task at hand but also guides the LLM to produce meaningful and contextually appropriate information. Thoughtful prompt construction is essential for harnessing the full potential of LLMs, ensuring that they generate outputs aligned with the user's intentions and the desired application \cite{pmlr-v202-shrivastava23a}. 

In addition, when using LLMs, a crucial parameter is the temperature, which influences the randomness and creativity of the model's output \cite{10.1162/tacl_a_00407}. The temperature parameter ranges from 0.0 to 1.0 and controls the probability distribution of the next word in a sequence generated by the model. A higher temperature, such as 1.0, introduces more randomness, allowing for diverse and imaginative responses. Conversely, a lower temperature, like 0.5, sharpens the focus of the model, leading to more deterministic and conservative outputs. Properly tuning the temperature is critical, as it directly impacts the balance between novelty and coherence in the generated text \cite{10.1145/3571730}.

\subsection{Project Management}\label{sec:PM} 

Project Management tools have been an integral part of the software engineering process since its dawn in the 1960s and 1970s, including tools like Microsoft Project\footurl{https://www.microsoft.com/es-es/microsoft-365/project}, which brought digitalization to project scheduling and resource allocation. 

With the advent of agile methodologies new tools were needed in order to tackle the fine-grained, high-volume information projects were generating. Platforms like Jira, Trello\footurl{https://trello.com/}, and Asana\footurl{https://asana.com/} emerged, emphasizing adaptability, real-time collaboration, and integration with other development tools. The significance of these platforms lies not only in task tracking, but also in their potential to foster communication, facilitate remote work, and adapt to the evolving needs of the software development industry. Jira has become one of the most widely-used Project Management tools, according to Google Trends\footurl{https://labur.eus/NNJxs}.

\section{JiraGPT Next}\label{sec:JiraGPT Next}

This section describes the design and implementation of JiraGPT Next, focusing on the user interface and the workflow to provide answers in natural language using GPT. Currently, it works as a web application compatible with any modern browser. It has been developed using Streamlit\footurl{https://streamlit.io/} as the library to build the web front-end and Python for back-end and every other task.

\begin{figure*}
    \centering
    \includegraphics[width=\textwidth] {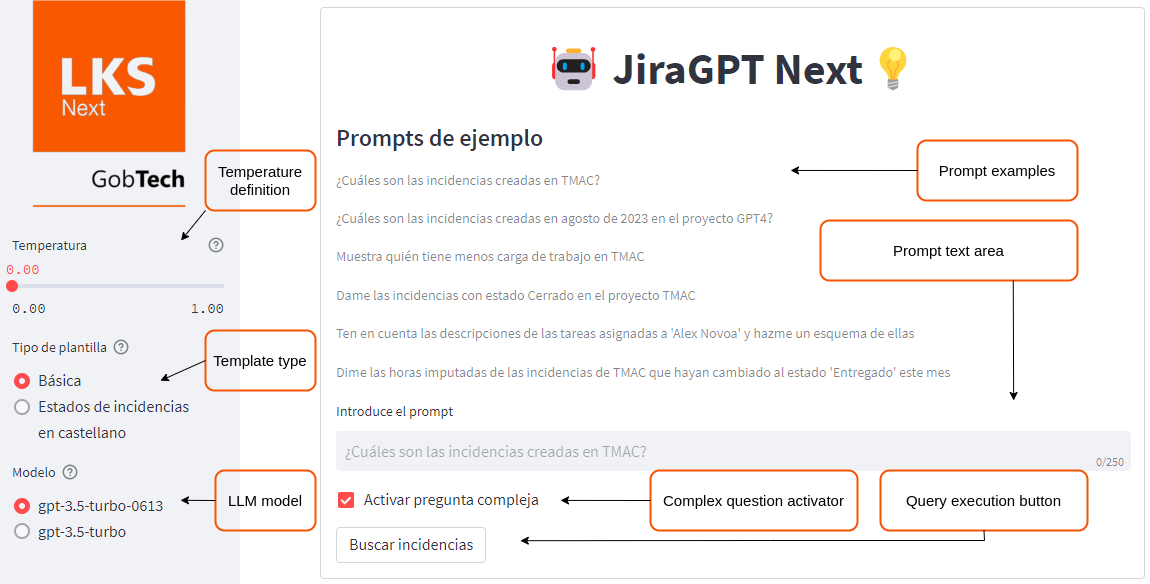}
    \caption{Screenshot of JiraGPT Next's user interface.}
    \label{fig:interface}
\end{figure*}

JiraGPT Next works as an add-on for a Jira installation and uses  Jira Query Language (JQL) to access its database. JQL is Jira's dedicated query language, which allows to construct queries based on specific criteria, such as issue types, statuses, assignees and custom fields. It enables advanced users and developers to precisely filter and retrieve information from the data contained in Jira. More information about JQL is available in its official documentation\footurl{https://www.atlassian.com/software/jira/guides/jql} and in the literature \cite{harned2018hands}.

\subsection{User Interface}

JiraGPT Next's user interface is presented in Figure \ref{fig:interface}. The  interface is divided in two panels, left and central, offering different functionalities. The central panel is designed to be used by project managers (end users of the application). In there, a project manager can type a query in natural language (e.g. ``How many issues have been were closed in January 2023?'') and the answer will be provided below. The panel is comprised of the following elements:

\begin{itemize}
    \item A list of examples as suggestions on how to type a query.
    \item The text area for the user's query.
    \item The option to mark the question as complex. More about this will be described later. 
    \item The button to submit the query.  
\end{itemize}

The panel in the left hand side is aimed to be used by developers and has been set for debugging purposes. It is currently used to test and tune the results of JiraGPT Next and, while it is visible in the current prototype, it will be hidden in future versions. It contains the following elements:

\begin{itemize}
    \item Temperature definition. Controls the temperature parameter of GPT, which leads to higher or lower randomness in the responses.
    \item Template type. 
    \item LLM model. It shows the available LLMs to use JiraGPT Next with.
\end{itemize}

We have defined that, in JiraGPT Next, a query made in natural language can be one of the following types:

\begin{itemize}
    \item Basic: A query whose answer is a list of items, e.g. the issues created in January 2023.
    \item Complex: A query whose answer goes beyond a list of items and shall be expressed in natural language, e.g. the number of projects with more than 7 ongoing issues.
\end{itemize}

\begin{figure*}
    \centering
    \includegraphics[width=\textwidth]{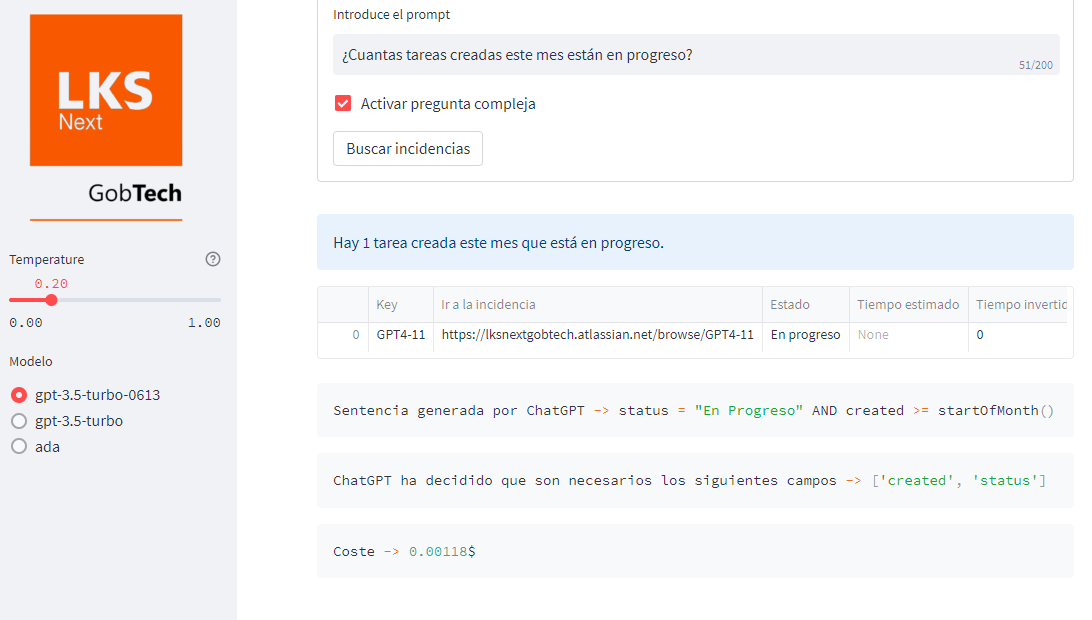}
    \caption{Screenshot of JiraGPT Next with the results to the query ``\textit{How many tasks created this month are in progress?}'' (In Spanish).}
    \label{fig:after-query}
\end{figure*}

In its current version, a user can define whether a query is basic or complex using the appropriate toggle in the user interface. The rationale behind this decision is to let the user have a way to control the cost linked to the queries. More about this will be detailed in the next section.

We show in Figure \ref{fig:after-query} a screenshot of JiraGPT Next with the results to a sample query ``How many tasks created this month are in progress?'', which is considered a complex query. JiraGPT Next provides an answer in natural language ``Hay 1 tarea creada este mes que está en progreso'' (``There's 1 task with status in progress created this month'') and a listing with information about this task, including a link that opens Jira with more information. Under this listing, the user finds the following information:

\begin{itemize}
    \item The JQL statement used to retrieve the information, which in the screenshot is: \texttt{status = 'En Progreso' AND created = startOfMonth()}.
    \item The Jira fields involved in the query.
    \item The cost of the query related to calls to the OpenAI GPT API.
\end{itemize}

\subsection{Query workflow}

\begin{figure*}
    \centering
    \includegraphics[width=\textwidth]{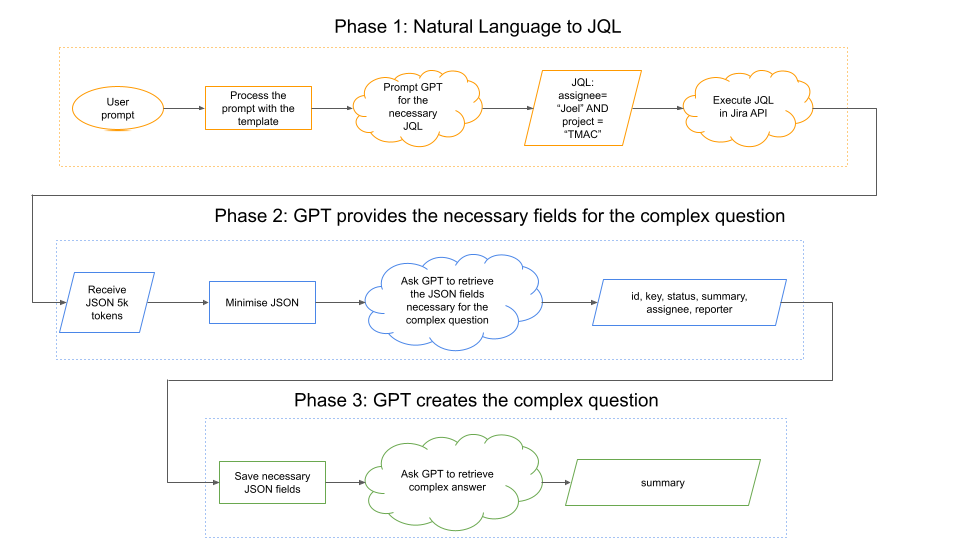}
    \caption{Phases of the JiraGPT Next internal process.}
    \label{fig:fases}
\end{figure*}

In general terms, JiraGPT Next takes the user's query, translates it to JQL, conducts some analysis and gives an answer to the end user. This process works in three phases, which have been depicted in Figure \ref{fig:fases}. Each phase requires a call to OpenAI's GPT API with a particular prompt, which will be described in depth in the next section. The phases of the workflow are:

\begin{itemize}
    \item Phase 1: The query of the user is coupled with a custom prompt and sent to GPT. GPT returns a query in JQL format and this query is executed in the local Jira instance.
    \item Phase 2: Results from the JQL query generated in the previous phase are collected and analyzed. The analysis retrieves the fields of the items in the results, couples them with a custom prompt and are sent to GPT. From the list of fields, GPT returns a subset of the fields required to construct an answer to the user in the next phase.
    \item Phase 3: The GPT API is used to construct an answer in natural language for the user. It requires using a custom prompt and subset of the data obtained in phase 1, guided by the field selection conducted in phase 2.
\end{itemize}

When a query is conducted as Basic, only the first phase of the workflow is used: JiraGPT Next generates a JQL query using the GPT API and runs it on the Jira instance. Jira returns a list of items corresponding to the JQL statement, and those are presented to the user. This mode only involves Phase 1 of the program, so it is the fastest and the least costly.

When a query is conducted as Complex (i.e. requires interpretations or calculations that can not be directly retrieved using JQL) the three phases of the workflow are used:

\begin{enumerate}
    \item The JQL is generated as in the Basic Mode.
    \item The JQL is executed and the Jira issues are received as a JSON file.
    \item In phase 2, fields from the JSON of the issues that are not useful for any type of question are removed. Only 21 fields are maintained for each issue.
    \item The GPT API is used to retrieve the fields in the JSON object that are strictly necessary to solve the question. GPT responds with a list of the fields and a new JSON is created with just those fields for each issue. The goal of this step is to simplify the input for the next phase.
    \item In phase 3, the reduced JSON and the user's question are send to GPT, which elaborates an answer in natural language. This answer is presented to the user in the web interface.
\end{enumerate}

\subsection{GPT Prompts}\label{sec:prompts}

In each phase of the workflow, a query is made to the GPT model. This query contains a prompt which serves as input for the model and is composed of a template and some input data (e.g. in phase 1, the query that the user types). This section describes the templates used in each phase.

\paragraph{\textbf{Phase 1 template}}

The aim of this phase is to generate the JQL query that will be executed in Jira. To this end, JiraGPT Next uses the following template prompt (the numbers next to each paragraph are references that will be used in Section \ref{sec:evaluation}; ``GPT4'' is the name of a sample project in Jira):

\begin{enumerate}
\item 
\begin{itquote}
You are an AI assistant trained on JIRA Query Language. Your task is to translate user requests into precise JQL queries. Remember, the output should only be the JQL query itself, without any additional information or explanations.
\end{itquote}
\item 
\begin{itquote}
Always use the following issue status names in Spanish and never in English: “Open” should be “Abierto”, “In Progress” should be “En Progreso”, “Resolved” should be “Resuelto”, “Approved” should be “Aprobada”, “Delivered” should be “Entregado”, “Reopened” should be “Reabierto”, “Closed” should be “Cerrado”.
\end{itquote}
\item 
\begin{itquote}
Please note: If a project name is not specified in the user’s request, do not invent or assume a project name. Simply omit the project name from the generated JQL query. For example, given the user’s query “Muestra las incidencias en progreso en GPT4”, you should return “status = “En Progreso” AND project = GPT4”’
\end{itquote}
\item 
\begin{itquote}
Another example: given the user’s query “¿Cuáles son las incidencias de máxima prioridad asignadas a joel.garcia?” you should return “assignee = “joel.garcia” AND priority = “Highest””.
\end{itquote}
\end{enumerate}

In this phase, if the query of the user was ``¿Cuántas personas tienen asignadas tareas en el proyecto GPT4?'' (``How many employees have tasks assigned in the project GPT4?''), GPT would return a JQL statement like \texttt{assignee is not empty AND project = GPT4}.

\paragraph{\textbf{Phase 2 template}}

In this phase, GPT is asked to select the JSON fields that are required to solve the query, among all of those returned by the JQL query. The following template prompt is used:

\begin{itquote}
Given the user’s query, please respond ONLY with the specific fields from the JIRA API’s JSON response, separated by commas, that are necessary to fulfill the user’s request. Each field should be separated by a comma with no additional explanation or text. For example, if the user’s request required the “assignee”, “project”, and “time” fields, you would respond: “assignee, project, time”. Now, please analyze the following user query
\end{itquote}

If the user query was ``¿Cuántas personas tienen asignadas tareas en el proyecto GPT4'' (``How many employees have tasks assigned in the project GPT4?''), after the first phase, this phase would return \texttt{Necessary JSON fields: [assignee]}.

\paragraph{\textbf{Phase 3 template}}

In this phase, GPT is asked to provide an answer in natural language. Appart from the template prompt, the user's query in natural language and the reduced JSON is provided. The template prompt is:

\begin{itquote}
You are part of an application that interfaces with JIRA. The program is as follows: A user’s natural language query asking about issues in JIRA is translated into Jira Query Language using GPT API. The JQL is then executed, and the resulting issues are collected in a JSON. You are given a JSON of the issues and the user’s query and you must answer with a response to the user’s query.
\end{itquote}

After this phase, if the user query was ``¿Cuántas personas tienen asignadas tareas en el proyecto GPT4'' (``How many employees have tasks assigned in the project GPT4?''), the answer would be \texttt{A partir del JSON de incidencias proporcionado, se puede determinar el número de personas diferentes a las que se les ha asignado una incidencia en GPT4. En este caso son 3 personas}. (``From the issues JSON provided, the number of different people who have been assigned an issue in GPT4 can be determined. In this case there are 3 people'').


\section{Evaluation}\label{sec:evaluation}

This section presents an evaluation of the accuracy of one of the prompts used in JiraGPT Next. Although the prompts in use have been described in section \ref{sec:prompts}, this section aims to give a detailed overview of the different options that have been considered during the design of the prompts and their impact on the effectiveness of the program.

We want to note that this section presents an evaluation solely of the prompt used in the first phase of the application for different reasons: firstly, the JQL generation stage is crucial for the correct functioning of JiraGPT Next and it can be evaluated accurately in an automatic way. Secondly, and with regards to the prompt of the second phase, even if it suggests more fields than the strictly necessary to create a query, JiraGPT Next would create correct results and the answer for the end user would not be affected. And thirdly, the assessment of the third phase is left out because, given that the answer is generated in natural language, its evaluation could be subjective.

For this evaluation, we prepared a set of questions that a project manager would type into JiraGPT Next. We compute the accuracy of a particular prompt by checking if the answer provided by GPT is right or not: for example, if a query asks for Jira issues in a particular period, we check JiraGPT Next's reply with the data in Jira. If it returns exactly the expected issues, the answer is marked as correct. If it does not return every issue expected by the user or returns some issues that should not be returned, the answer is marked as incorrect. We do not evaluate whether the generated JQL query matches an already established one because GPT can generate a large number of JQL statements that are correct but not considered by us. As a final note, these tests have used GPT v3.5 and were run between September and October 2023.

\subsection{Environmental setup}\label{sec:env-setup}

We created a Jira test project with 2 users and 20 issues. We consider that an issue can be in one of the following states: Open (\textit{``Abierto''}), In progress (\textit{``En progreso''}), Solved (\textit{``Resuelto''}), Validated (\textit{``Validado''}), Handed (\textit{``Entregado''}), Closed (\textit{``Cerrado''}), Reopen (\textit{``Reabierto''}). In this project, 14 issues were in ``Open'' state , 1 was ``In progress'', 1 was ``Solved'', 1 was ``Validated'', 1 was ``Handed'', 1 was ``Closed'' and 1 was ``Reopen''.

In addition, we prepared a set of 70 test questions that a project manager would type in JiraGPT Next, based on the experience of LKS Next staff. We consider that each question belongs to one of the following categories:
\begin{itemize}
    \item Type 1: Questions whose answer can be retrieved filtering issues by one field, e.g. \textit{List every issue created in august 2023}, which requires using the \textit{date} field.
    \item Type 2: Questions whose answer can be retrieved filtering issues by two or more fields, e.g. \textit{List every issue that was on approved state in august 2023}, which requires using the \textit{date} and \textit{status} field.
    \item Type 3: Questions that require additional interpretation to provide an answer, e.g. \textit{How many issues have changed to In Progress in august 2023?}.
\end{itemize}

Among the 70 questions, 34\% of them were of type 1, 34\% of type 2 and 32\% of type 3, in order to provide an even distribution in the dataset.

\subsection{Results}\label{sec:results}

These tests will show the effects of using different prompts to convert a natural language query into a JQL query that, once executed, returns a set of issues. The prompts that will be tested are the text blocks that form the prompt of the first phase, presented in section \ref{sec:prompts}. These have been tested incrementally, with the first test using Block 1 as a prompt, the second test using Blocks 1 and 2 as prompt, and so on.

These prompts can be categorized as zero-shot or few-shot. Zero-shot prompts are those that give a task to the LLM without any specific examples or instructions, whereas few-shot prompts provide the LLM with a limited set of examples or instructions relevant to the task. In essence, the difference between zero-shot and few-shot prompting lies in the level of guidance provided to the LLM. Zero-shot relies on the model's innate language understanding, while few-shot offers a limited but task-specific context to enhance performance. More information about this can be found in the literature \cite{zhou2023large}.

For each prompt we report its accuracy and token cost. The former represents the number of queries to which has provided a correct answer, out of the 70 in the test set. The latter represents the amount of tokens required by GPT to process the query as a measure of cost. In GPT, one token corresponds roughly to 4 characters of text for common English text. Token usage has been calculated using OpenAI's Tokenizer\footurl{https://platform.openai.com/tokenizer}.

The first test used the first block as prompt, which obtained an accuracy of 17,14\%.  This prompt can be considered as zero-shot because it does not provide any example or further guidance beyond the request, which can explain its low accuracy. An analysis of the results revealed that this prompt did not generate a correct JQL for any question that required querying the status of a Jira issue. It also failed in every question that did not provide explicitly the name of the Jira project in use, because GPT made up names that did not match with that of the test project. This prompt uses 44 tokens.

The second test used the first and second block as prompt, which obtained an accuracy of 22,86\%. This prompt results in an improvement in the accuracy compared to the previous one mainly due to the inclusion of status names. The issues in the test project have status names in Spanish and GPT assumes those to be English, hence questions that failed for this reason with the previous prompt are correct with this one. This prompt uses 136 tokens.

The third test used the first, second and third blocks as prompt, which obtained an accuracy of 37,14\%. The increase in accuracy compared to the previous prompt can be explained by three reasons: the first one is that the inclusion of the third block fixes a previously mentioned problem: GPT inventing the name of the project when it is not provided as part of the user query. The second one is the inclusion of a positive and a negative request in the prompt. LLM-related literature \cite{10.1145/3544548.3581388} suggests that this practice improves the quality of the results, as it receives an example of what to do and what to not do. In this case \textit{...Omit the project name...} is considered a positive request and \textit{...do not invent or assume...} is considered a negative request. The third one is that providing an example makes this prompt of few-shot type, which has been proved to be more effective \cite{zhou2023large}. This prompt uses 219 tokens.

The fourth and last test considered the full template used for the first stage of JiraGPT Next as prompt, which obtained an accuracy of 48,57\%. The inclusion of a second example as part of the prompt has resulted in an increase of the accuracy, mainly because the number of questions related to priority correctly answered is higher compared to the previous prompt. This prompt uses 272 tokens.

To summarize these tests, we provide a comparison of the results in Table \ref{tab:sum-results}. We can observe how, the more complete the prompt, the better the accuracy - but this comes with an increase in the token usage. As an example, accuracy is 2.8 times better using the full prompt compared to using just the first block, but this comes with the cost of using 6.1 times more tokens. Including more examples to the prompt would probably improve its accuracy but incurring in higher costs.

\begin{table}[hbt]
\centering
\begin{tabular}{|r|r|r|}
\hline
\multicolumn{1}{|l|}{\textbf{Prompt blocks}} & \multicolumn{1}{l|}{\textbf{Accuracy}} & \multicolumn{1}{l|}{\textbf{Required tokens}} \\ \hline
1                                            & 17,14\%                                & 44                                            \\ \hline
1 and 2                                         & 22,86\%                                & 136                                           \\ \hline
1, 2 and 3                                      & 37,14\%                                & 219                                           \\ \hline
Full                                         & 48,57\%                                & 272                                           \\ \hline
\end{tabular}
\caption{Summary of results for the tests with different prompts.}
\label{tab:sum-results}
\end{table}

\begin{figure}[hbt]
    \centering
    \includegraphics[width=\linewidth,trim={0.5cm 0 0.4cm 0}]{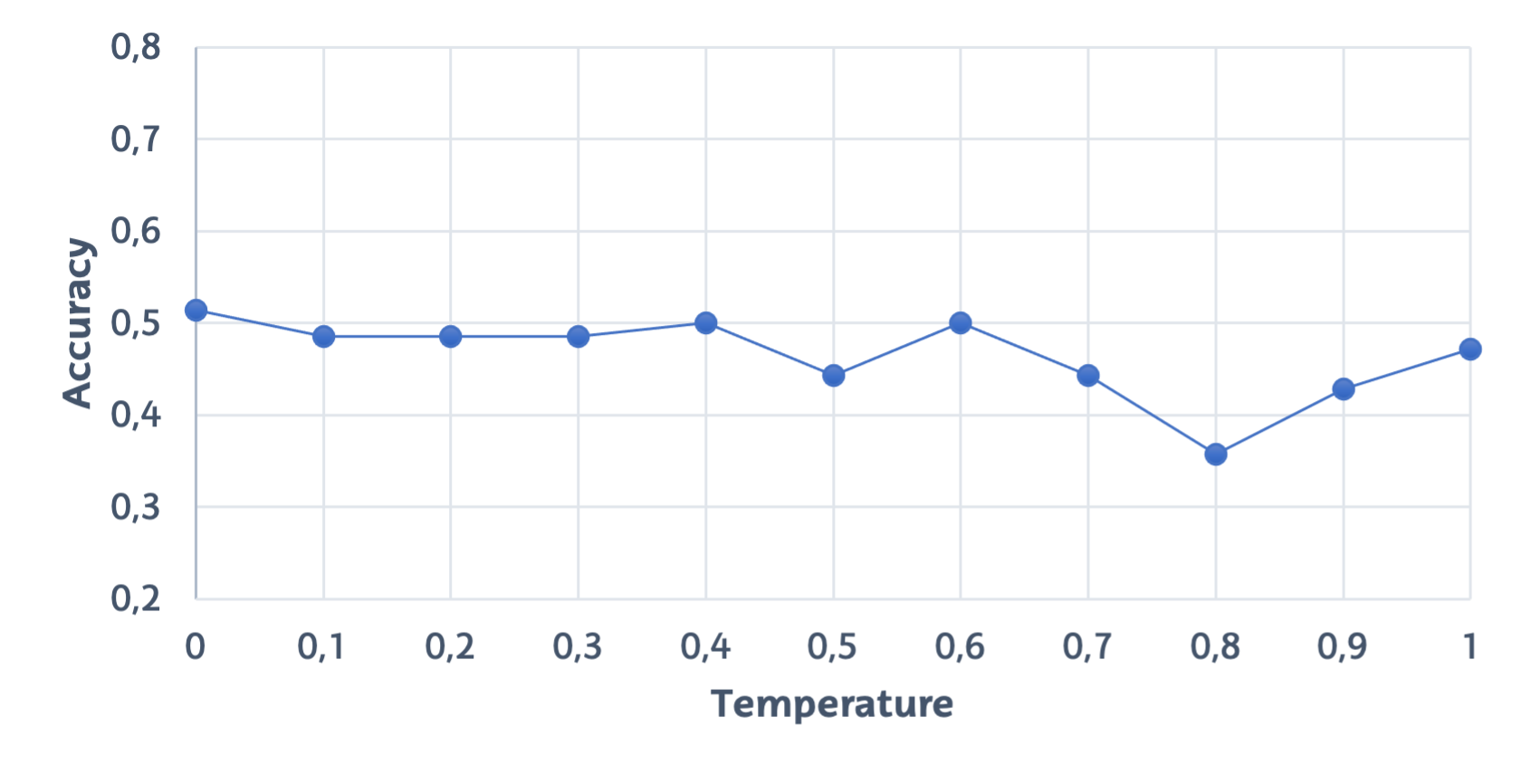}
    \caption{Accuracy of the full prompt with different temperature values.}
    \label{fig:temperature}
\end{figure}

Finally, we assessed the effect of the temperature on the results of GPT. We run the same test using the full prompt but with different temperature values, from 0 to 1 with 0.1 steps. Results are shown in Figure \ref{fig:temperature}. We can observe how 0 temperature provides the best results and 0.8 the worst ones. We believe that this is due to the nature of the task requested to GPT: the generation of a JQL query is a demand for a structured text where no creativity or randomness is required. Given that lower temperature values produce more deterministic outputs, this could explain why values lower than 0.5 produce the best results.

\section{Related products}\label{sec:related-products} 

JiraGPT Next is not the only software using AI to improve Project Management tasks. We conducted a review of the existing ones and present in this section a selection of the most relevant three: Kubiya.ai, Microsoft 365 Copilot, and Albus. 

Kubiya.ai\footurl{https://www.kubiya.ai} is a virtual assistant developed as a Slack\footurl{https://slack.com/} bot. Its main building blocks are Actions and Workflows: Actions are API calls that retrieve, create or modify data (e.g. to the GitHub API). Workflows are flows of Actions that Kubiya.ai decides to use based on what the user has written as a prompt. Workflows are composed of Actions and a user can configure both. At the time of writing, the cost is \$40 per month/user and there is no integration with tools other than Slack like Jira (Only 10 Actions are provided), GitLab, Nexus, Sonarqube or Teams\footnote{These are in use at LKS Next, but are also widely used in the industry.}.

Microsoft 365 Copilot\footurl{https://adoption.microsoft.com/copilot} is an assistant integrated into Microsoft Teams, offering help to users of the Microsoft 365 Applications suite. It includes a plugin for the integration with Jira. At the time of writing, the license cost is 30\$ per month/user and there are per-region established regulatory constraints.

Albus\footurl{https://www.springworks.in/albus} is an AI platform that can integrate services like JIRA, Drive and Dropbox with a virtual assistant hosted at Slack. Even though it offers integrations with Jira, it lacks features like including links to Jira issues in the response. At the time of writing, each user account costs 10\$/month with a limit of 100 questions/month.

The development and evaluation of JiraGPT Next spanned over three months and used 493.955 tokens of the GPT API, which incurred in the cost of \textasciitilde{}1\$. Even if the business model is not defined yet, we expect the usage cost of JiraGPT Next to be significantly lower than the described products.


\section{Conclusions}\label{sec:concl}

This work has presented JiraGPT Next, a software that extends Jira's functionality providing the ability to make requests in natural language. At its core lies a process that uses the GPT LLM to firstly transform user requests from text to JQL, then makes some filtering and finally provides a reply back to the end user in natural language. We described its user interface and presented an evaluation of the accuracy and cost of several prompts used as input for GPT.

The motivation of this work is to ease the way that project managers and team members interact with their Jira installation. This capability does not only promise to streamline Project Management practices but also to democratize the use of Jira, making it more accessible to those without extensive prior training or technical expertise. 

Future work will mainly span in two directions. One will be the improvement of the LLM-based process that produces the answer to users' queries. As described in the Evaluation, an increase in the length of the prompt implies more costs, so we will consider the inclusion of more examples or the tunning of existing ones. We will also consider using other LLMs like Google's Gemini\footurl{https://gemini.google.com} and Meta's Llama\footurl{https://ai.meta.com/llama}, which have proved to be solid alternatives to GPT. We will also explore the possibility of adding a RAG (Retrieval Augmented Generation) component to improve the answers of the system.

The other will be the improvement of JiraGPT Next as a product. Currently, the software is considered to be a prototytpe which requires further testing and tuning. Next steps will involve the  improvement and addition of more features, plus conducting tests with a large user base to assess the quality of its replies with queries from different people and receive feedback from its usability.

\section*{\uppercase{Acknowledgements}}

This work was partially supported by the grant GAITECH: \textit{Cognitive assistance supported by GenAI for software management and development in the Public Administration} (ZL-2024/00685), funded by the Basque Government.








\bibliographystyle{apalike}
{\small
\bibliography{Example}}



\end{document}